\documentclass{article}%
\usepackage{amsmath}
\usepackage{amsfonts}
\usepackage{amssymb}
\usepackage{graphicx}
\usepackage{epstopdf}%
\providecommand{\U}[1]{\protect\rule{.1in}{.1in}}

\ifx\pdfoutput\relax\let\pdfoutput=\undefined\fi
\newcount\msipdfoutput
\ifx\pdfoutput\undefined\else
\ifcase\pdfoutput\else
\ifx\paperwidth\undefined\else
\ifdim\paperheight=0pt\relax\else\pdfpageheight\paperheight\fi
\ifdim\paperwidth=0pt\relax\else\pdfpagewidth\paperwidth\fi
\fi\fi\fi
\begin{document}

\title{A criticism of "gas mode "reinterpretations of the Michelson-Morley and
similar experiments.}
\author{Daniel Shanahan\\18 Sommersea Drive, Raby Bay, Queensland \ 4163, Australia}
\date{}
\maketitle

\begin{abstract}
It has been argued by R. T. Cahill and others that a Michelson interferometer
in "gas mode" - in which the light paths are through an included gaseous
medium - are able to detect and have detected an absolute frame of
reference.\ \ It is shown here that this argument supposes incorrectly that
the reduced velocity of light in gas is an observer-independent invariant.
\ This velocity is not invariant, but given in a frame with respect to which
the interferometer moves with velocity $v$ by the usual relativistic formula
for the addition of velocities, these being in this case the velocity $v$ and
the reduced velocity of light in the inertial frame of the interferometer.
\ It is suggested nonetheless that though the absolute frame urged by Cahill
may be undetectable, there are persuasive grounds for considering the
alternative Lorentzian Relativity that did suppose the existence of such a
frame.\medskip\medskip\medskip\medskip

(This is an updating of a note that appeared in the March/April 2006 issue of
\textit{Australian Physics} \cite{shanahan}. \ A similar criticism of Cahill's
"gas mode" analysis has been given by Sfarti \cite{sfarti}.)

\end{abstract}

\medskip\medskip

R. T. Cahill has argued that the Michelson-Morley experiment of 1887
\cite{michelson} did not produce the null result generally supposed (see
Cahill and Kitto \cite{cahillkitto2002}, and Cahill \cite{cahill2003b} to
\cite{cahill2010}). \ This note draws attention to an apparent flaw in that
argument, and in similar proposals by Consoli et al \cite{consoli2} to
\cite{consoli4} and Demjanov \cite{demjanov1} \cite{demjanov2}. \ However, it
is also suggested here that whether or not the absolute frame urged by these
investigators is detectable, there exist compelling reasons for considering
the alternative Lorentzian approach to relativity that did suppose such a frame.

Fringe shifts were observed by Michelson and Morley, but these were much
smaller than those expected from the presumed velocity of the interferometer
through the aether. \ Cahill contends that these shifts were consistent
nonetheless with the passage through an absolute frame of reference of an
interferometer in "gas mode" - an interferometer in which the light paths are
not through vacuum but through contained air or other gaseous medium. \ 

The null result reported by Michelson and Morley has been confirmed by later
experiments of greater precision (see, for instance, Refs. \cite{braxmaier} to
\cite{hermann}). \ Moreover, the modern view of these experiments, or at least
the prevailing modern view, is that any aether or other absolute frame of
reference is likely to remain undetected because of the covariance of the
Lorentz Transformation (LT). \ That seems to have been the conclusion reached
eventually by Lorentz himself \cite{lorentz}, as also by Poincar\'{e} who was
the first to describe the group theoretic properties underlying that
covariance \cite{poincare}.

Cahill accepts that such a frame will not be revealed by an interferometer in
which the light paths are through vacuum. \ In that case, differences in
travel time will be countered exactly by Fitzgerald-Lorentz contraction. \ He
argues, however, that where, as in Michelson-Morley and similar experiments,
the light paths are through contained air, the balancing is not exact.
\ Because of the diminished phase velocity of light in gas, small phase shifts
will occur of the order of magnitude of those reported. \ 

This "gas mode" analysis and the theory of "process\ physics" to which it has
led \cite{cahill2003b} have gained some support in the literature (see Refs.
\cite{consoli2} to \cite{consoli4}, and \cite{eckhardt}). \ Indeed, Cahill's
reasoning has been extended to the "vacuum mode" by Consoli et al by relying
on vacuum fluctuations and gravitationally induced refraction effects
\cite{consoli2} to \cite{consoli4}. \ Cahill has himself proposed that noise
in interference fringe readings could be evidence of gravity waves
\cite{cahill2006} to \cite{cahill2010}.

These proposals assume incorrectly that, like the velocity $c$ of light in
vacuum, the reduced phase velocity $V$ of light through gas is an invariant.
\ It is only to an observer in the inertial frame of the interferometer that
the refracted light has the particular velocity that has been supposed by
Cahill. \ In considering the magnitude of this velocity from another inertial
frame, one must apply the relativistic formula for the addition of velocities,
those velocities being in this case, the velocity of the interferometer with
respect to the other frame, and the phase velocity $V$ of the signal with
respect to the gas entrapped within the interferometer. (See Shanahan
\cite{shanahan}, and Sfarti \cite{sfarti}).\ \ 

The relativistic addition formula, given by Einstein in 1905 \cite{einstein1},
was applied by von Laue as early as 1907 in calculating the velocity of light
in a moving medium \cite{laue}. \ It is perhaps ironic that, as recounted
recently by Janssen \cite{janssen}, the moving medium was in that case the
aether - \ von Laue applied the addition formula in calculating the effect of
Fresnel drag. \ 

As Janssen has also mentioned,\ von Laue's calculation was subsequently cited
by Einstein in a text on relativity published in Germany in 1917
\cite{einstein2}. \ It is also discussed in Pauli's well-known work on
relativity \cite{pauli}, at p. 17, while the application of the addition
formula to the velocity of light in a moving medium is routinely mentioned in
modern texts on optics and relativity, for instance, Resnick \cite{resnick},
at p. 80, Rindler \cite{rindler}, at p. 44, and Jenkins and White
\cite{jenkins}, at p 421.

Let us suppose (see Fig. 1) that the Michelson interferometer is moving with
velocity $v$ with respect to some absolute frame of reference, and that it is
only in this frame that light is not only observed to have, but in fact has,
the velocity $c$ in vacuum. \ Where the interferometer arm is aligned in
parallel with the direction of $v$, the formula for the addition of velocities gives,%

\begin{equation}
V_{A_{1}B}=\frac{V+v}{1+Vv/c^{2}}, \label{vab}%
\end{equation}
for the forward path, and%
\begin{equation}
V_{BA_{2}}=\frac{V-v}{1-Vv/c^{2}}, \label{vba}%
\end{equation}
for the return path. \ 

As observed from the absolute frame, the arm length $L$ is relativistically
contracted to $L/\gamma$, where $\gamma$ is the usual Lorentz factor,%
\[
(1-\frac{v^{2}}{c^{2}})^{-1/2}
\]
so that the time for a round trip\ as considered from that frame becomes,
\begin{align*}
t_{A_{1}BA_{2}}  &  =\frac{L/\gamma}{(V+v)/\left(  1+Vv/c^{2}\right)
-v}+\frac{L/\gamma}{(V-v)/\left(  1-Vv/c^{2}\right)  +v}\\
& \\
&  =2\gamma L/V
\end{align*}

In the other, the light paths are inclined at an angle to the direction of
travel. \ \ However we need only consider those components of phase velocity
and path length normal to the direction of travel (or see equivalently Sfarti
\cite{sfarti}, Eqn. (2.6) et seq.). \ The time-dilated phase velocity is
$V/\gamma$, while the corresponding component of path length is simply $L$.
\ Thus,%
\[
t_{A_{1}C}=t_{CA_{2}}=\gamma L/V,
\]
from which,%
\[
t_{A_{1}BA_{2}}-t_{A_{1}C\,A_{2}}=0,
\]
for all $V$, ensuring a null result whatever the gas within the interferometer.

But instead of the relativistically composed velocities (\ref{vab}) and
(\ref{vba}), Cahill takes their simple Galilean equivalents, that is,%
\[
V_{A_{1}B}=V+v,
\]
for the forward path, and%
\[
V_{BA_{2}}=V-v,
\]
\ for the return path. \ (See, for example, the discussion preceding Eqn. (16)
in Cahill \cite{cahill2006}, at p. 79). \ \ 

Proceeding in this way,\ he finds, after some algebra, a difference in travel
times,
\[
t_{A_{1}BA_{2}}-t_{A_{1}CA_{2}}=\frac{(n^{2}-1)L}{c}\frac{v^{2}}{c^{2}%
}+O(\frac{v^{4}}{c^{4}})+..,
\]
from which he naturally, but incorrectly, concludes that a Michelson
interferometer in gas mode may act as a detector of absolute motion.

The assumption that the phase velocity $V$ is invariant invites the further
criticism that the same effective index of refraction for the contained gas
has been assumed throughout the experiment, whatever the orientation and
velocity of the apparatus. \ Refractive index is a function of density, and
from the standpoint of an observer in the absolute frame, the density of the
gas encountered by the light signal will depend on whether it is moving with
or against or transversely to the movement of the gas.

However, the null results reported from such experiments do not have today the
significance they had in 1887. \ A clearly established \textit{positive}
result would compel a reconsideration of the invariability of the laws of
physics and presumably some surprise that Nature is not quite as elegant as
the covariance of the LT would suggest. \ But, as noticed above, a null result
does not itself preclude the existence of an aether or other absolute frame of
reference (or theories of gravity that presuppose such an aether). \ 

Several writers (notably Bell \cite{bell}, but see also, for example, Miller
\cite{miller} and Nelson \cite{nelson}) have stressed the pedagogical merits
of studying not only the Special Relativity of Einstein, to whom the existence
of an aether was "superfluous", but also the earlier approach of Lorentz,
Larmor and Poincar\'{e}, who did suppose a preferred frame and
aether.\ \ Briefly stated, the distinction between Einstein's theory and what
is commonly called Lorentzian Relativity is that the former explains the LT as
a physical transformation of spacetime and the latter from changes in the
structure of matter as it suffers a boost from one inertial frame to another
(see generally Shanahan \cite{shanahan2}). \ 

Following the publication of his famous paper of 1905 \cite{einstein1},
Einstein's theory eventually displaced the older Lorentzian approach.
\ Lorentz had struggled to explain why all aspects of Nature transform in like
manner, a problem compounded by the apparent inconsistency of Lorentz's
essentially classical electromagnetic modelling with the new quantum
mechanics. \ His explanations of the various changes described by the LT also
seemed piecemeal and ad hoc - lacking in the kind of unifying principle
supplied by Einsteins "postulate" that the laws of physics must appear the
same for all observers, a principle that was to find elegant expression in
Minkowski's geometrical treatment of spacetime \cite{minkowski}. \ And of
course special relativity was naturally seen as precursor to Einstein's
success with general relativity (GR).

Yet for all its elegance and economy, there is something elusive in Einstein's
theory. \ There seems little doubt that the LT is an accurate description of
the changes of length, time and simultaneity \textit{observed} on a change of
inertial frame. \ But the notion that the LT describes some actual (as
distinct from observed) change in space really makes no sense at all. \ No
such difficulty arises in GR where a change in the metric affects in like
manner all that occupies the part of space in question. \ But how are we to
comprehend that space is able to contract in one way for one particle and in a
different way for another moving relatively to the first, albeit that the two
(or at least their correspondingly contracted fields) could be occupying the
very same piece of space? \ And are we to suppose that when an astronaut
observes the contraction of the constellations as she accelerates through
space, she really believes that the stars are closing ranks about her?

As the astronaut suffers a boost from one inertial frame to another, the only
relevant physical changes actually occurring are in her and her spacecraft.
\ Neither the stars nor the space occupied by the stars can have changed in
any relevant way. \ It must be some change in the astronaut that not only
causes her to appear transformed to others, but induces in that astronaut the
illusion that it is not she and her space craft that has changed, but all else
about her.\ \ This then is the Lorentzian view of things - the explanation of
the LT as a consequence of changes in matter as it changes inertial frame.\ \ 

On this view, it becomes necessary to look beyond the symmetry of the metric
and ask questions that are precluded in Einstein's theory. \ Why are the laws
of physics the same for all observers? \ And what after all is the origin of
the observed metric? \ If the LT describes not actual, but merely perceived,
changes in space and time, the invariance of the laws of physics cannot
provide the unifying principle suggested by special relativity. \ Nor can the
Minkowski metric be the source of that invariance for it must itself be seen
as a kind of illusion induced by changes occurring in the observer. \ 

The proposition that the metric is not given but emergent - that it
supervenes\ on matter - has been argued forcefully (see Brown \cite{brown} and
Brown and Pooley \cite{brownpooley2}), and just as strongly resisted (see, for
instance, Janssen \cite{janssen} \cite{janssen2}, Norton \cite{norton} and
Nerlich \cite{nerlich}). \ But that resistance has been largely grounded on
the simplicity and elegance of Einstein's theory. \ Once it is realized that
the LT describes perceived rather than actual changes in space (or spacetime),
the LT, the metric, and the invariance of physical laws, must all be seen as
having their origin in some even more fundamental property of Nature. \ 

A hint of a deeper unifying principle was given in 1925 when de Broglie
proposed that matter shares the wave-like characteristics of light
\cite{debroglie}, thereby uncovering, as Einstein put it, "a corner of the
great veil". \ It then became possible to argue that everything in Nature
transforms in like manner because it is constituted in like manner from the
same underlying wave-like influences.

It can be shown that the merits of such an Lorentzian approach are far more
than pedagogical. \ A compelling advantage is the emergence of the otherwise
anomalous de Broglie wave, not as the independent wave generally supposed, but
as a modulation defining the dephasing of the underlying wave structure in the
direction of travel (as seems to have been first noticed in a different
context by R. Horodecki \cite{horodecki}, see also Macken \cite{macken}, and
other references in Shanahan \cite{shanahan2}). \ 

Considered as a modulation, the superluminal velocity\ of the de Broglie wave
is no longer that of energy transport and need not be explained away by the
awkward device of recovering the classical velocity of the particle from the
group velocity of a packet of such de Broglie waves. \ This interpretation of
the de Broglie wave sheds light in turn on the role of the wave in
quantization and in the Schr\"{o}dinger and Dirac Equations, and explains the
relevance of the de Broglie wave number to the\ optical properties of massive
particles. \ The dephasing defined by the modulation becomes the underlying
physical explanation of the failure of simultaneity that is perhaps the most
counter-intuitive aspect of SR. \ 

These matters are discussed in more detail in Ref. \cite{shanahan2}.


\begin{thebibliography}{99}                                                                                               %


\bibitem {shanahan}\ \ D. Shanahan, \ \textit{Australian Physics, }Mar./Apr. (2006)

\bibitem {sfarti}\ \ A. Sfarti, Corrected theory of the reenactments of the
Michelson-Morley experiment in non-vacuum media, Rom. J. Phys. \textbf{52},
533 (2007)

\bibitem {michelson}\ A. A. Michelson and E. W. Morley, On the Relative Motion
of the Earth and the Luminiferous ether, Am. J. Sc. \textbf{34}, 333 (1887)

\bibitem {cahillkitto2002}\ R. T. Cahill, and K. Kitto, Michelson-Morley
Experiments Revisited and the Cosmic Background Radiation Preferred Frame.
Apeiron \textbf{10}, 104 (2003)

\bibitem {cahill2003b}\ R. T. Cahill, \textit{Process Physics: From
Information Theory to Quantum Space and Matter, }Nova Science, New York (2005)\ 

\bibitem {cahill2005b}\ R. T. Cahill, The Michelson and Morley 1887 Experiment
and the Discovery of Absolute Motion, Prog. Phys. \textbf{3}, 25 (2005)

\bibitem {cahillletter}\ R. T. Cahill, \textit{Australian Physics, }Jan./Feb. (2006)

\bibitem {cahill2006}\ \ R. T. Cahill, A New Light-Speed Anisotropy
Experiment:\ Absolute Motion and Gravitational Waves Detected, Prog. Phys.
\textbf{4}, 73 (2006)

\bibitem {cahill2007}\ \ R. T. Cahill, Dynamical 3-Space: A Review, in M.
Duffy and J. L\'{e}vy eds., \textit{Ether space-time and cosmology: New
insights into a key physical medium, }Apeiron (2009)

\bibitem {cahill2008}\ R. T. Cahill, Resolving Spacecraft Earth-Flyby
Anomalies with Measured Light Speed Anisotropy, Prog. Phys. \textbf{4}, 9 (2008)

\bibitem {cahill2008b}\ R. T. Cahill, Unravelling Lorentz Covariance and the
Spacetime Formalism, Prog. Phys. \textbf{4}, 19 (2008)

\bibitem {cahill2009}\ R. T. Cahill, Combining NASA/JPL One-way Optical-fibre
Light-Speed Data with Spacecraft Earth-Flyby Doppler-Shift Data to
Characterize 3-Space Flow, Prog. Phys. \textbf{4}, 50 (2009)

\bibitem {cahill2010}R. T. Cahill, Lunar Laser-Ranging Detection of Light
Speed Anisotropy and Gravitational Waves, arXiv: 1001.2358 [physics.gen-ph] (2010)

\bibitem {consoli2}\ M. Consoli, A. Pagano, and L. Pappalardo, Phys. Lett. A
\textbf{318}, 292\ (2003)

\bibitem {consoli3}\ M. Consoli and E. Costanzo, The motion of the Solar
System and the Michelson-Morley experiment, arXiv: astro-ph/0311576 (2003)

\bibitem {consoli}\ M. Consoli and E. Costanzo, From classical to modern
ether-drift experiments: the narrow window for a preferred frame, Phys. Lett.
A, \textbf{333}, 355 (2004)

\bibitem {consoli4}\ M. Consoli and E. Costanzo, Old and new ether-drift
experiments: a sharp test for a preferred frame, N. Cimento B119, 393 (2004)

\bibitem {demjanov1}\ \ V. V. Demjanov, Physical interpretation of the fringe
shift measured on Michelson interferometer in optical media, Phys. Lett. A
\ \textbf{374}, 1110 (2010)

\bibitem {demjanov2}\ \ V. V. Demjanov, What and how the Michelson
interferometer measures, arXiv:1003.2899v6 [phys.gen-ph] (2010)

\bibitem {braxmaier}\ C. Braxmaier, H. M\"{u}ller, O. Pradl, J. Mlynek, A.
Peters, and S. Schiller, Tests of Relativity Using a Cryogenic Optical
Resonator, Phys. Rev. Lett. \textbf{88}, 010401 (2002)

\bibitem {wolf}\ \ P. Wolf, S. Bize, A. Clairon, A. N. Luiten, G. Santarelli,
and M. E. Tobar, Tests of Lorentz Invariance using a Microwave Resonator,
Phys. Rev. Lett. \textbf{90}, 060402 (2003)

\bibitem {lipa}\ \ J. A. Lipa, J. A. Nissen, S. Wang, D. A. Stricker, and D.
Avaloff, New Limit on Signals of Lorentz Violation in Electrodynamics", Phys.
Rev. Lett. \textbf{90}, 060403 (2003).

\bibitem {muller}\ H. M\"{u}ller, P. L. Stanwix, M. E. Tobar, E. Ivanov, P.
Wolf, S. Herrmann, A. Senger, E. V. Kovalchuk, and A. Peters, "Relativity
tests by complementary rotating Michelson--Morley experiments". Phys. Rev.
Lett. \textbf{99,} 050401 (2007)

\bibitem {eisele}\ C. Eisele, A. Y. Nevsky, and S. Schiller, Laboratory test
of the isotropy of light propagation at the 10-17 level, Phys. Rev. Lett.
\textbf{103}, 090401 (2009)

\bibitem {hermann}\ S. Herrmann, A. Senger, K. M\"{o}hle, M. Nagel, E. V.
Kovalchuk, and A. Peters, Rotating optical cavity experiment testing Lorentz
invariance at the 10-17 level, Phys. Rev. D \textbf{80, }105011 (2009)\ 

\bibitem {lorentz}\ H. A. Lorentz, \textit{The Theory of Electrons}, Teubner,
Leipzig (1916)

\bibitem {poincare}\ H. Poincar\'{e}, Sur la dynamique de l'\'{e}lectron,
Rendiconti del Circolo matematico di Palermo \textbf{21}, 129 (1906)

\bibitem {eckhardt}\ \ H. Eckhardt, An Alternative Hypothesis for Special
Relativity, Prog. Phys. \textbf{2}, 56 (2009)

\bibitem {einstein1}\ A. Einstein, Zur elektrodynamik bewegter korper, Ann.
Phys. \textbf{17}, 891 (1905). English trans., On the electrodynamics of
moving bodies, in H. A. Lorentz, A. Einstein, H. Minkowski, H. Weyl,
\textit{The Principle of Relativity}, Methuen, London (1923)

\bibitem {laue}\ M. von Laue, Die Mitf\"{u}hrung des Lichtes durch bewegte
K\"{o}rper nach dem Relativit\"{a}tsprinzip, Ann. Phys. \textbf{23}, 989 (1907)

\bibitem {janssen}\ M. Janssen, Drawing the line between kinematics and
dynamics in special relativity, Studies in History and Philosophy of Modern
Physics \textbf{40}, 26 (2009)

\bibitem {einstein2}\ A. Einstein, \textit{\"{U}ber die spezielle und
allgemeine Relativit\"{a}tstheorie (gemeinverst\"{a}ndlich)}, Vieweg,
Braunschweig (1920), English trans., \textit{Relativity}, Crown, New York (1961)

\bibitem {pauli}\ W. Pauli, \textit{Theory of Relativity}, Dover, New York (1958)

\bibitem {resnick}\ R. Resnick, \textit{Introduction to Special Relativity,
}Wiley, New York (1968)

\bibitem {rindler}\ W. Rindler, \textit{Introduction to Special Relativity},
Oxford, New York (1982)

\bibitem {jenkins}\ F. A. Jenkins and H. A. White, \textit{Fundamentals of
Optics}, 4th Ed. McGraw-Hill, Singapore (1981)

\bibitem {bell}\ J. S. Bell, How to teach Special Relativity, Prog. Sci. Cult.
1, 2 (1976), reprinted in: \textit{Speakable and Unspeakable in Quantum
Mechanics}, revised edn., Cambridge University Press, Cambridge (2004)

\bibitem {miller}\ D. J. Miller, A constructive approach to the special theory
of relativity, Am. J. Phys. \textbf{78}, 633 (2010)

\bibitem {nelson}\ W. M. Nelson, A wave-centric view of special relativity,
arXiv:1305.3022 physics.class-ph (2013)

\bibitem {shanahan2}\ \ D. Shanahan, A Case for Lorentzian Relativity, Found.
Phys. \textbf{44}, 349 (2014) (DOI 10.1007/s10701-013-9765-x)

\bibitem {minkowski}H. Minkowski, Raum und Zeit, Phys. Zeits. \textbf{10}, 104
(1909), English trans.: Space and time, in H. A. Lorentz, A. Einstein, H.
Minkowski, H. Weyl, \textit{The Principle of Relativity}, Methuen, London (1923)

\bibitem {brown}\ H. S. Brown, \textit{Physical Relativity},\textit{\ }Oxford
University Press,\textit{\ }Oxford (2005)

\bibitem {brownpooley2}H. S. Brown and O. Pooley, Minkowski space-time: a
glorious non-entity, in D. Dieks (ed.), \textit{The Ontology of Spacetime,}
Elsevier, Amsterdam (2006)

\bibitem {janssen2}M. Janssen, Reconsidering a Scientific Revolution: The Case
of Einstein \textit{versus} Lorentz, Phys. Perspect. \textbf{4}, 421 (2002)

\bibitem {norton}\ J. D. Norton, Why Constructive Relativity Fails, Brit. J.
Phil. Sci. \textbf{59}, 821(2008)

\bibitem {nerlich}\ G. Nerlich, Bell's 'Lorentzian Pedagogy': A Bad Education
[preprint] philsci-archive.pitt.edu/5454/1/Bell.pdf (2010)

\bibitem {debroglie}\ L. de Broglie, Ph. D. Thesis, Recherches sur la
th\'{e}orie des quanta. Ann. de Phys. (10) \textbf{3}, 22 (1925). English
trans., Researches on the quantum theory, in Ann. Fond. Louis de Broglie
\textbf{17}, 92 (1992)

\bibitem {horodecki}\ R. Horodecki, Information Concept of the Aether and its
application in the Relativistic Wave Mechanics and Quantum Cybernetics, in L.
Kostro, A. Poslewnik, J. Pykacz, and M. \.{Z}ukowski, Eds., \textit{Problems
in Quantum Physics, Gdansk '87}, World Scientific, Singapore (1987)

\bibitem {macken}\ J. A. Macken, \textit{The Universe is Only Spacetime}, http://onlyspacetime.com/
\end{thebibliography}
\end{document}